\documentclass[conference]{IEEEtran}
\IEEEoverridecommandlockouts
\usepackage{amsmath,amssymb,amsfonts}
\usepackage{algorithmic}
\usepackage{graphicx}
\usepackage{textcomp}
\usepackage{xcolor}

\usepackage{algorithm}
\usepackage{algorithmic}
\usepackage{svg}
\svgsetup{inkscapelatex=false}
\usepackage[style=ieee, sorting=none]{biblatex}
\def\BibTeX{{\rm B\kern-.05em{\sc i\kern-.025em b}\kern-.08em
    T\kern-.1667em\lower.7ex\hbox{E}\kern-.125emX}}
\IEEEoverridecommandlockouts\IEEEpubid{\makebox[\columnwidth]{ 978-1-6654-2849-1/21/\$31.00~\copyright~2021 IEEE \hfill} \hspace{\columnsep}\makebox[\columnwidth]{ }}
    
\begin{document}

\title{Wireless Sensor Networks for Optimisation of Search and Rescue Management in Floods}

\makeatletter
\newcommand{\linebreakand}{%
  \end{@IEEEauthorhalign}
  \hfill\mbox{}\par
  \mbox{}\hfill\begin{@IEEEauthorhalign}
}
\makeatother

\author{
  \IEEEauthorblockN{Harshil Bhatt}
  \IEEEauthorblockA{\textit{Department of Electronics} \\
    \textit{and Communication Engineering}\\
    \textit{Manipal Institute of Technology} \\
    Manipal, India \\
    harshilbhatt2001@gmail.com}
  \and
  \IEEEauthorblockN{Pranesh G}
  \IEEEauthorblockA{\textit{Department of Mechatronics} \\
   \textit{Engineering}\\
    \textit{Manipal Institute of Technology}\\
    Manipal, India \\
    praneshgopaln@gmail.com}
  \and
  \IEEEauthorblockN{Samarth Shankar}
  \IEEEauthorblockA{\textit{Department of Electronics} \\
    \textit{and Communication Engineering}\\
    \textit{Manipal Institute of Technology} \\
    Manipal, India \\
    samshankar18@gmail.com}
  \linebreakand 
  \IEEEauthorblockN{Shriyash Haralikar}
  \IEEEauthorblockA{\textit{Department of Electronics} \\
  \textit{and Instrumentation Engineering} \\
    \textit{Manipal Institute of Technology }\\
    Manipal, India \\
    shriyashharlikar30@gmail.com}
 
}

\maketitle
\begin{abstract}
We propose a novel search-and-rescue management method that relies on the aerial deployment of Wireless Sensor Network (WSN) for locating victims after floods. The sensor nodes will collect vital information such as heat signatures for detecting human presence and location, the flow of flood. The sensor modules are packed in a portable floating buoy with a user interface to convey emergency messages to the base station. 
Sensor nodes are designed based on disaster conditions, cost-effectiveness and deployed in the affected region by a centrifugal dispersion system from a helicopter. 
 A mobile ad-hoc network is set up by modifying the Low Energy Adaptive Cluster Hierarchy (LEACH) protocol for greater efficiency and adoption of multi-hop of Cluster Heads for long-distance communication to Base Station.
The model metrics have been defined considering previous rural floods in India. The efficiency and power characteristics of the network are are compared to other protocols via simulations.
The sensor data from the network makes resource management, rescue planning and emergency priority more efficient, thus saving more lives from floods.
\end{abstract}

\begin{IEEEkeywords}
Wireless Sensor Network, Disaster Management, Search and Rescue operation, Energy efficient networks
\end{IEEEkeywords}

\section{Introduction}
Floods are the worst natural disaster that claims more lives and casualties than any other disasters in the world. Every year, an average of 1700 people are killed, and 39 lakh people are affected by the ravages of the flood in India \cite{Tripathi2015}.
 The frequency of massive floods has increased due to climate change and global warming. Recurrent floods have led to high causalities and huge economic loss to the country. Due to floods, the estimated loss for the country is close to 1.5 per cent of GDP in 1915-2015 \cite{Panwar2020}.
Rural floods are more devastating since most of rural population live close to rivers.
The poor infrastructure and remote location have been a critical challenge in disaster management in rural areas. 
With the development of WSN, their application in disaster management has received significant attention since they can collect data more accurately with minimum human intervention. 
The WSN can help in collection of on-ground information and effectively coordinate the rescue operation. 
WSNs are resilient and remain online even during infrastructural collapse due to disaster unlike their wired counterparts. In addition to resilience, they are also are easy to scale and configure according to the requirement.
The previous research on WSN in disaster management has generally focused on the ease of deployment \cite{Jahir2019} and utilizing UAV to increase the surveillance \cite{Akbas2011}.
To effectively harness the full potential of WSNs, one needs to consider the main challenges faced by WSN for disaster management.

The primary challenge with WSN during a disaster are 

\textbf{\textit{Battery source and management:}} WSNs consume much more power to transceive data than wired counterparts. The power consumption increases with the amount of data it sends. 
Also, a vast number of nodes increases the complexity of power management, decreasing the network's efficiency.

\textbf{\textit{Network Architecture:}} It plays a dominant role in the complexity and effective routing of the data to the base station. The simple architecture is challenging to scale and complex architecture are computationally expensive. Also, the regular protocol does not account for the efficient power consumption of the node. There is a trade-off between computational complexity and efficiency in data transmission in the selection of network architecture.

\textbf{\textit{Sensors:}} Since there is a limitation on the quantity of data transmitted, sensors need to be carefully selected by trading off between the nodes' resolution and power efficiency. Besides the restriction to the amount of data, the minimum number of sensors needs to be determined to get the required data to keep the cost low and increase deployment scale.

\textbf{\textit{Sensor Node deployment:}}  The deployment needs to ensure uniform distribution, cover the complete area of disaster, ensure safe landing and retrieval of sensor nodes. The mode of deployment is also crucial since this will directly affect the time taken for deployment and survival rate of each node. 

\textbf{\textit{Complexity:}} The time taken for the relief to arrive once WSNs are deployed crucial. Management of a large number of sensor nodes and data is a difficult task. 

Considering these challenges, We propose a solution based on deployment of WSN during the time of flood to acquire real-time data from the affected region that could help the response team act better.
The primary contribution of this paper are
\begin{itemize}

\item Designed WSN and sensor node for flood relief application in rural areas with minimum infrastructure or damaged infrastructure. The design ensures scalability, cost-effectiveness, ease of deployment, and retrieval of the sensor node.
\item Proposed a novel Modified LEACH (MLEACH) protocol for effective routing to ensure minimum loss of data and speedy delivery of data packets to a base station.
\item Proposed a layered method with the abstraction of data between each layer to decrease complexity and improve data management in the network.
\item Proposed a unique method to filter out irrelevant and stationary data within the cluster and thus preventing the data traffic at the base station.

\end{itemize}
Thus, we propose power and cost-effective WSN for transmitting vital on-ground data like the victim's location, the flow of flood, and the priority of emergency will better optimize the rescue operation with limited response time and resources.

\begin{figure}[h]
  \centering
  \includegraphics[width=3in]{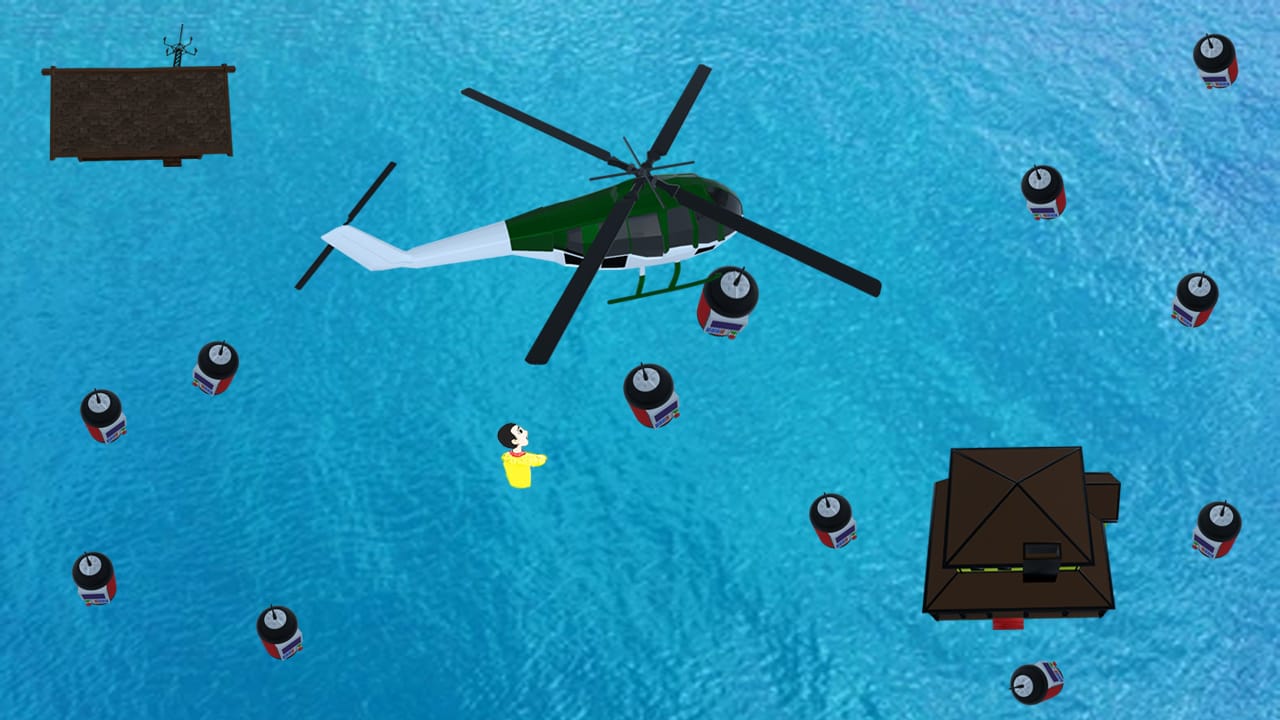}
  \caption{The WSN deployment for collecting data of the victims stuck in flood for optimizing the relief operation}
  \label{fig:improvedall_in_one}
\end{figure}

Rest of the paper is organised as follows: Section \ref{section:Literature_Survey} summarises the relevant works followed by section \ref{section:Sensor_Node_Specification}, discusses the deployment and node specifications. Subsequently, Section \ref{section:Communication} summarises the Modified LEACH (MLEACH) protocol. Section \ref{section:Simulation} puts forth the simulation results along with the comparison of proposed Communication model against traditional model and is concluded with Section \ref{section:Conclusion} and Section \ref{section:Acknowledgment}.

\section{Literature Survey}
\label{section:Literature_Survey}
Our work builds on crucial previous work in WSN and remote sensing. 
Interest in Early Warning Systems and Remote Sensing has picked up in the last few years. Balis et al. used WSNs for real-time remote sensing for Early Warning Systems and presented a framework to facilitate early warning systems \cite{Balis1}. Mojaddadi et al. using remote sensing data and Geographic Information System proposed an ensemble method to create flood probability indices \cite{Mojaddadi1}. These models are resource-intensive as they use satellite data which generates large amounts of data. Supercomputers are used to manage and work on the data to predict floods. These resources are not feasible for under-developed or developing countries.

A significant challenge is the lifetime of the WSN, as it should last for the entire duration of the search and rescue operation. Since energy is a scarce resource that limits the lifetime of WSNs, several research works have proposed energy-efficient routing protocols. Heinzelman et al. proposed Low Energy Adaptive Clustering Hierarchy (LEACH). This clustering-based protocol utilises the randomised rotation of local cluster heads to distribute energy load among sensors nodes in the network evenly. The significant drawbacks of this protocol were the single-hop communication between cluster heads and the Base Station, and remaining energy in sensor nodes was not considered during the election of cluster heads \cite{Heinzelman1}. Sabet et al. proposed an energy-efficient decentralised hierarchical cluster-based multi-hop routing algorithm for WSNs. In their approach, clustering and multi-hop routing algorithms perform simultaneously to reduce the energy consumption due to control packets \cite{Sabet1}. Singh et al. compared different variations of the LEACH routing protocol based on parameters like clustering, overhead, scalability, load balancing, and complexity for WSNs \cite{Singh1}. Kodali et al. presented a multi-level hierarchical protocol based on LEACH protocol which improved the energy efficiency and lifetime of the network \cite{Kodali1}. Since the cluster head (CH) in LEACH is random, the probability of a node becoming a CH is equal for a node with low energy and high energy. This affects the robustness and lifetime of the network. Since LEACH follows single-hop communication between the CH and Base Station (BS), the energy required by CH increases after a certain distance to transmit data to the BS. This leads to uneven dissipation, where nodes far from the BS die quickly; thus, reducing the effective sensing area. These issues are addressed by implementing multi-hop communication between the CH and BS along with beneficial attributes of existing LEACH protocol. Also, the proposed model improves the random selection of CH by preventing the same node from becoming a CH for the next $n$ rounds. 

Survivor detection during disaster management is a popular topic amongst researchers, and several sensors and algorithms have been proposed. Portmann et al. proposed a tracking algorithm based on particle filter combined with background subtraction for detecting and tracking people in thermal-infrared images. \cite{Portmann1}. Byunghun et al. analyse the performance and applicability of passive infrared motion sensors in security systems and propose a region-based human tracking algorithm \cite{Byunghun1}. While these algorithms substantially increase human detection accuracy and detection speed, it is computationally expensive to deploy on each sensor node. The advantage of using passive infrared sensors is the reduction in processing power and cost.

Aerial deployment is one of the easiest, effective and robust way to deploy WSN during a disaster.  The Deployment method needs to achieve maximum coverage in a minimum time. Sharma et al. proposed a Centrifugal Cannon based Sprinkler random deployment algorithm to scatter sensor nodes from a helicopter moving over the candidate region. This model is time efficient and scatters over vast and unreachable areas while minimising the number of scans over the area \cite{Sharma1}. In contrast, random dispersal mechanism may cause coverage holes right below the vehicle, thus making targeted dropping nearly impossible.

\begin{figure}[h]
  \centering
  \includegraphics[width=3.5in]{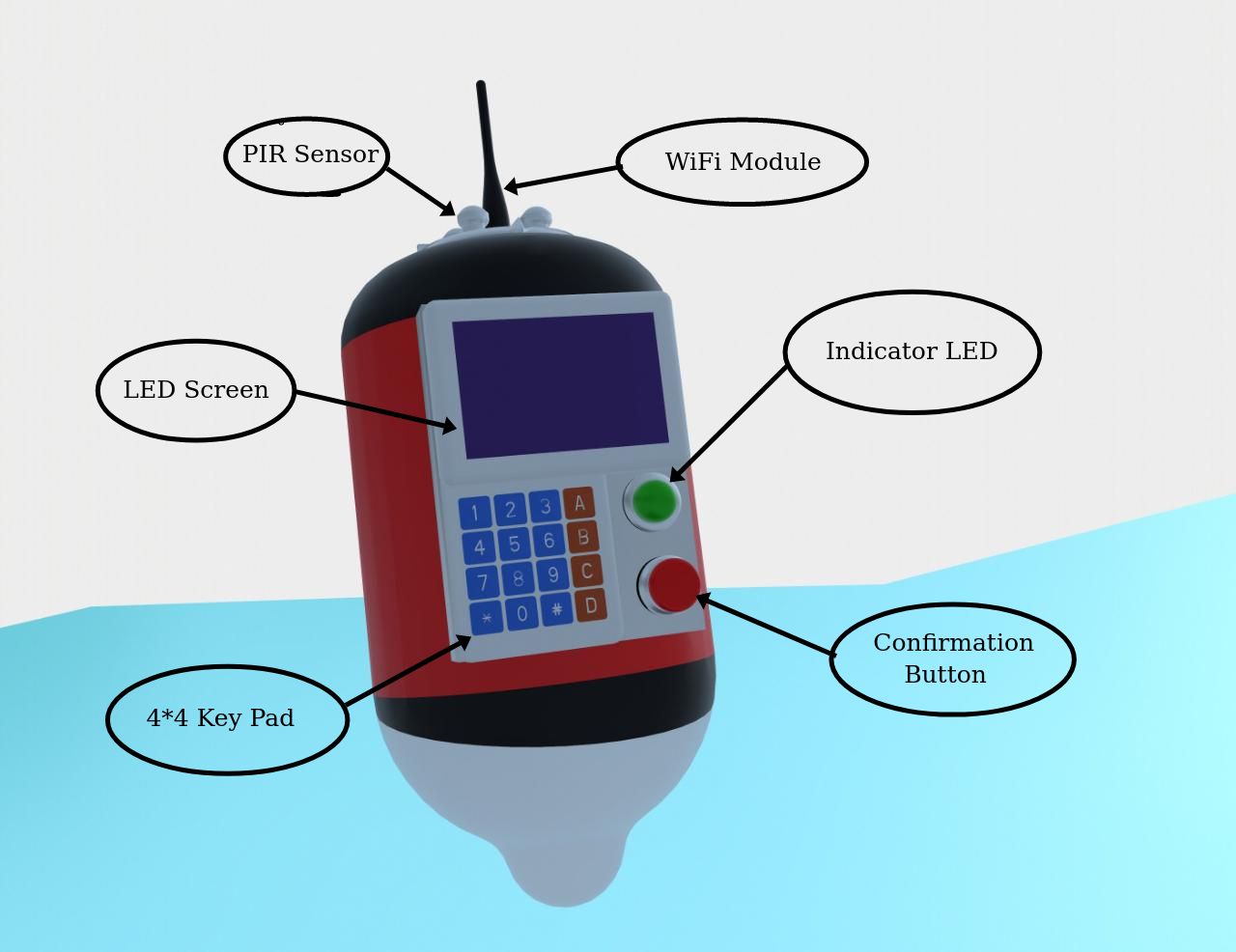}
  \caption{CAD model of the Sensor Node}
  \label{fig:cadmodel}
\end{figure}

\section{Sensor Node Specification}
\label{section:Sensor_Node_Specification}

The design of the proposed senor node is presented with utmost considerations of both the safety and robustness aspect of the node. The overview of the proposed solution is given by fig \ref{fig:improvedall_in_one} The whole section consists of three parts, Design and Deployment, Hardware Architecture as mention in \ref{fig:node_specs} and Software Architecture.

\subsection{Design and Deployment}
After a Base Station is setup at a suitable location, the deployment of the WSN is carried out by the helicopter with a centrifugal canon sprinkler mechanism with added perpendicular drop system to scatter the nodes in all direction and in significantly less time \cite{Sharma1}. The pod proposed shown in fig \ref{fig:cadmodel} is made out of PVC foam with a density of $0.5 g/cm^3$ to ensure floating and cushions in landing during aerial deployment. The pod is designed to provide stability in water by ensuring that the centre of gravity of the sensor node, at no point is above the metacentre by placing heavy components close to the base.  The designed pod is 45 cm in height, and the maximum diameter of the body is 30 cm. The whole pod with all the electronics weigh less than a kilogram, making it a light and mobile sensor node so that victims can carry it to a safe place in the locality to wait for the response team.  GPS and the activity status light would prove beneficial in retrieving sensor nodes after use for multiple deployments.  

\begin{figure}
    \centering
    \includegraphics[width=3.7in]{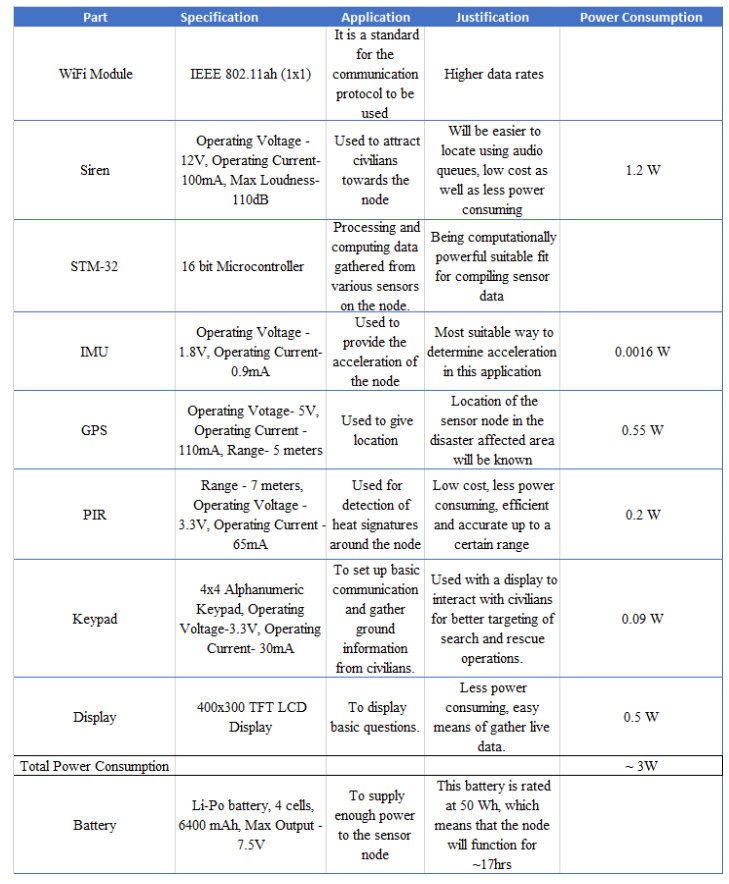}
    \caption{Sensor Node Specifications}
    \label{fig:node_specs}
\end{figure}

 \subsection{Software Architecture}

After the initial deployment of the WSN, only the sensors, siren, and WiFi module are in an active state to save energy. Siren is programmed to create loud noise at certain intervals to attract people. The operational status can be inferred from the green LED, and the push button is placed along with it to ensure the human presence with utmost certainty. An LCD screen and keypad are installed on the sensor node to receive on-ground information from the victim regarding the number of victims trapped, the severity of injury, age group, and the presence of any medical emergency. 
The microcontroller has multiple pipelines for polling sensor data, transceiving data across nodes and processing data. These are ensured by multiple threads in the controller and switching between the task in high frequency. A Real-Time Operating System (RTOS) is used to ease scheduling and decrease latency. The RTOS also helps set priority levels for different tasks, where human detection by the PIR sensor can be prioritized more. 
A set of predetermined weights are associated based on the level of emergency according to the data from the victims to optimise the rescue operation with limited resources. With on-ground data, a weighted graph optimisation problem can be formulated as a familiar Travelling Salesman Problem(TSP) to find both effective and efficient routes for relief operations.

\section{Communication}
\label{section:Communication}
\subsection{Energy Radio model}
The communication protocol has been designed to focus on the energy efficiency of the networks. 
A dynamic energy model is considered, where the energy dissipation depends on the number of packets being transmitted. Here, the transmitter and receiver of the wireless sensor node spend 50nJ/bit energy.
We consider the transmitter of the amplifier requires $\epsilon_{amp} = 120 PJ /bit/m^2$ to get the required range of Signal to Noise Ratio (SNR).

\begin{align}
& E_{Tx}(k,d) = E_{Tx-elec}(k) + E_{Tx-amp}(k,d) \\
& E_{Tx}(k,d) = E_{elec}*k + \epsilon_{amp}*k*d^2 \\
& E_{Rx}(k) = E_{Rx-elec}(k)\\
& E_{Rx}(k) = E_{elec}*k 
\end{align}

where,\\
$E_{elec}$ = Minimum \; electronics\; energy required to run Tx or Rx circuitry, \\
$E_{Tx-elec}$  = Energy consumed due to transmit electronics, \\
$E_{Rx-elec}$  = Energy consumed due to receive electronics, \\
$ \epsilon_{amp} $   = Transmit Amplifier energy consumption, \\
K   – number of message bits, \\
d - distance in meters,\\
$E_{Tx}$  = Total Transmit energy consumed,\\
$E_{Rx}$  = Total Receive energy consumed, \\

Equation (2) and (4) give the total energy expended by transmitter and receiver for n bits message and d meters distance. Thus, we get the total energy consumed by wireless sensor nodes \cite{Kodali1}. We use this radio energy model in our MLEACH  protocol.

\subsection{Communication Protocol}
The communication protocol proposed in this paper is an energy-efficient Modified LEACH (MLEACH) protocol. 
The nodes in LEACH group themselves into local clusters, with one node serving as cluster-head (CH).The usage of clusters for data transmission minimizes the effective transmit distance between transceivers by implementing hops among CH necessary to reach the base station. 

LEACH is a single hop routing protocol where each cluster head transmits data directly to the base station. The limitation of a node to communicate only in the vicinity of the base station and  consumes significant amount of energy for long distance communication making it unsuitable for disaster management applications. 
To surmount this problem we propose MLEACH protocol which is a multi-hop and layered protocol, making it apt for search and rescue operation. In addition to this, an unique data filtering method is introduced to reduce the data congestion and increase the energy efficiency of the networks.

The functioning of the protocol is separated into two phases: setup and data transmission.

\textbf{Set-up Phase:}
In this phase,The CH election, cluster formation and assignment of the Time Division Multiple Access (TDMA) slots to each Cluster Member (CM) is done. 
All nodes participate in the CH selection step by creating arbitrary precedence value in 0-1 range. If a sensor node's produced arbitrary number is smaller than a threshold value $T(n)$, the node gets elected as CH. The value of T(n) is calculated using equation (5) \cite{Heinzelman1}.

\begin{equation}
T(n) = 
\begin{cases}
\frac{P}{1-P*(r mod\frac{1}{P})}  : if n \in G 
\\
0 : else
\end{cases}
\label{eqn:cluster_head_election}
\end{equation}

$P$ represents the percentage of node to emerge as CH among nodes, $r$ is the ongoing round, and $G$ is collection of all nodes that were excluded in the CH selection in $1/P$ rounds. This equation ensures that any node that hasn't been a CH has an equal probability of becoming one and energy gets distributed equitably among the sensor nodes \cite{Singh1}. Once the CH is elected, it broadcasts 'hello' messages to the nearby nodes and adds them to the cluster as CM. After the cluster formation, the CH creates a TDMA and allocates a time slot to each CM to transmit the sensor data to the CH. 
The TDMA slots prevents CM data from colliding and allows their transmitter to be switched off until their slot is available. This lowers each node's energy usage and extends the network's lifespan.

\textbf{Transmission Phase:}
The network is divided into the lower and upper layers for transmission of data to the base station. 
\textit{Lower Layer:}
This layer consists of all the CM's. It deals with a conglomeration of sensor data from the CM on the selected CH.
\textit{Higher Layer:}
This layer consists of all the CH's.
Here, all the CH's processes the collected data from CM and filter the sensor data if a CM has not changed beyond the threshold. Thus, only essential data is transmitted to the base station via Dijkstra's algorithm through other CH to achieve the shortest route, ultimately reducing data traffic and improving energy efficiency.  
 \begin{algorithm}
    \caption{Proposed Routing Algorithm}
    \label{alg1}
    \begin{algorithmic}[1]
        \STATE begin round
        \IF{node!=CH}
            \WHILE{TDMA Slot exists}
                \IF {Node has data}
                \STATE Transmit Data $\rightarrow$ CH
                \ELSE 
                \STATE Send Heart Beat message $\rightarrow$ CH
                \ENDIF
            \ENDWHILE
                
        \ELSE
        \STATE Broadcast schedule $\rightarrow$ CM's and receive data from them
            \IF{Received data $>$ Thresholds}
            \STATE Send hello packet $\rightarrow$ neighbouring CH
            \STATE Populate Routing Tables of each CH 
            \STATE initiate Dijkstra's algorithm and send data $\rightarrow$ BS
            \STATE End Round
            \ELSE
            \STATE Delete Data Packet 
            \ENDIF
        \ENDIF
    \end{algorithmic}
\end{algorithm}

\begin{figure}
    \centering
    \includegraphics[width=3.5in]{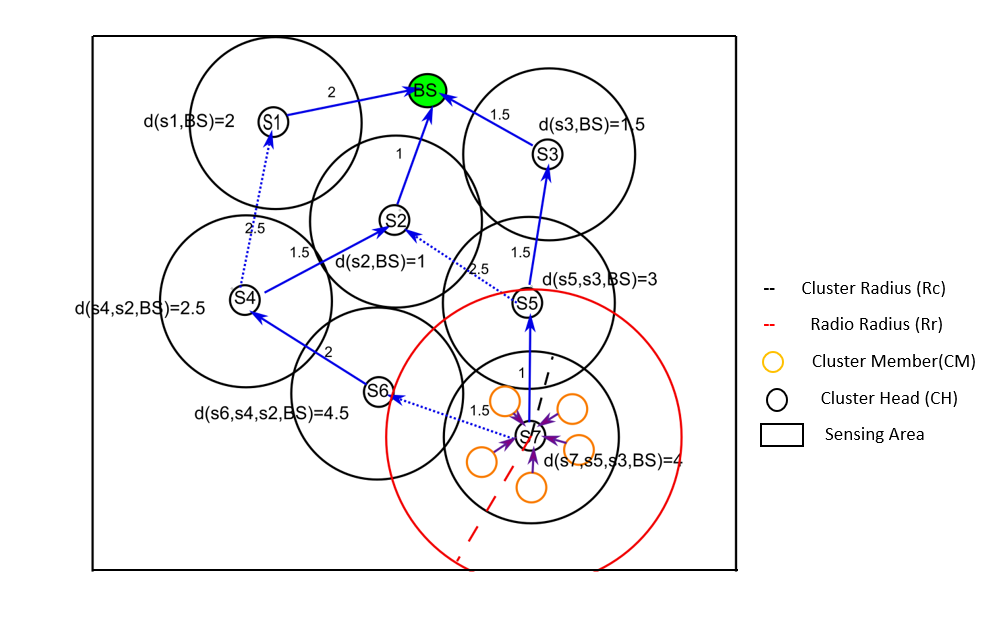}
    \caption{Routing Tree Diagram}
    \label{fig:routing}
\end{figure}
In figure \ref{fig:routing}, we have illustrated the construction of the routing tree.
Lower layer data transmission:
All the cluster members of cluster head S1, S2, S3, S4, S5, S6, and S7 form the lower level. In the figure in the cluster radius (Rc) of the S7 cluster head, there are 5 cluster members. They transmit data to S7 according to their TDMA slot. 
Higher layer data transmission:
All the cluster heads S1, S2, S3, S4, S5, S6, and S7 form the higher layer.
The base station is in the radio range(Rr) of S1, S2,  and S3  so, the data is transmitted directly to the BS from them. For S4 CH there are two routes $S4 -> S2 -> BS \; and \; S4 -> S1 -> BS$ . As $d(S4,S1,BS)=4.5 > d(S4,S2,BS)= 2.5$ , the shortest route $S4 -> S2 -> BS$ is chosen for transmission of data. Similarly, the routes are determined for S5, S6, and S7.

\section{Simulation}
\label{section:Simulation}
In this section, we evaluate the performance of our protocol and compare it with another pro-active routing algorithm, Destination-Sequenced Distance Vector (DSDV) via simulations. The simulation platform used is Network Simulator 3 (NS-3). We assume that the BS is fixed and lies at a random location within the sensor field. All the sensor nodes are randomly and non-uniformly distributed in the sensor field. A random way-point mobility model is used for simulating the movement of the sensor nodes. The IEEE 802.11ah wireless networking protocol is used to provide internet connectivity to the nodes. An On-Off Application was used to generate traffic to the BS in the network. The simulation parameters are summarized in Table \ref{table:parameters_simulation}.

\begin{table}[h!]
\centering
\begin{tabular}{|c|c|} 
 \hline
 \textbf{Parameters}                 & \textbf{Values}     \\ [0.5ex] 
 \hline
 Number of Nodes            & 512                          \\ 
 \hline
 Sensor Field               & 7500m*7500m                  \\
 \hline
 BS Location                & Random (within sensor field) \\
 \hline
 Data Packet Size           & 512 bytes                    \\
 \hline
 Initial Energy of nodes    & 172800 J                     \\
 \hline
 Average Energy Consumption & 0.836 J                  \\
 \hline
 Simulation Time            & 120s                         \\ [1ex] 
 \hline
\end{tabular}
\caption{Parameters of Simulation}
\label{table:parameters_simulation}
\end{table}

\begin{figure}[]
    \centering
    \includegraphics[width=3.5in]{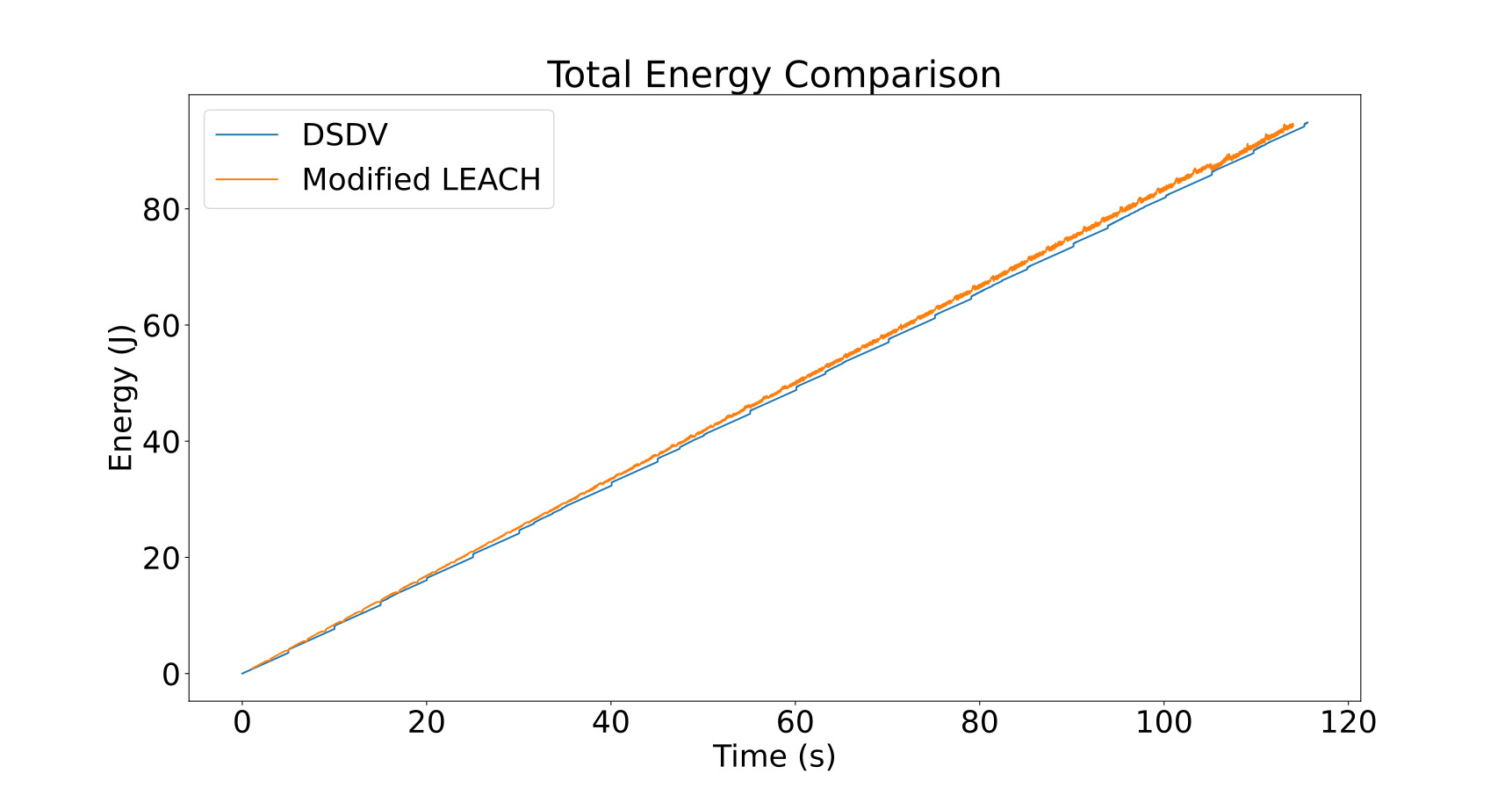}
    \caption{Total Energy vs Time of DSDV and MLEACH}
    \label{fig:total_energy}
\end{figure}

\begin{figure}[]
    \centering
    \includegraphics[width=3.5in]{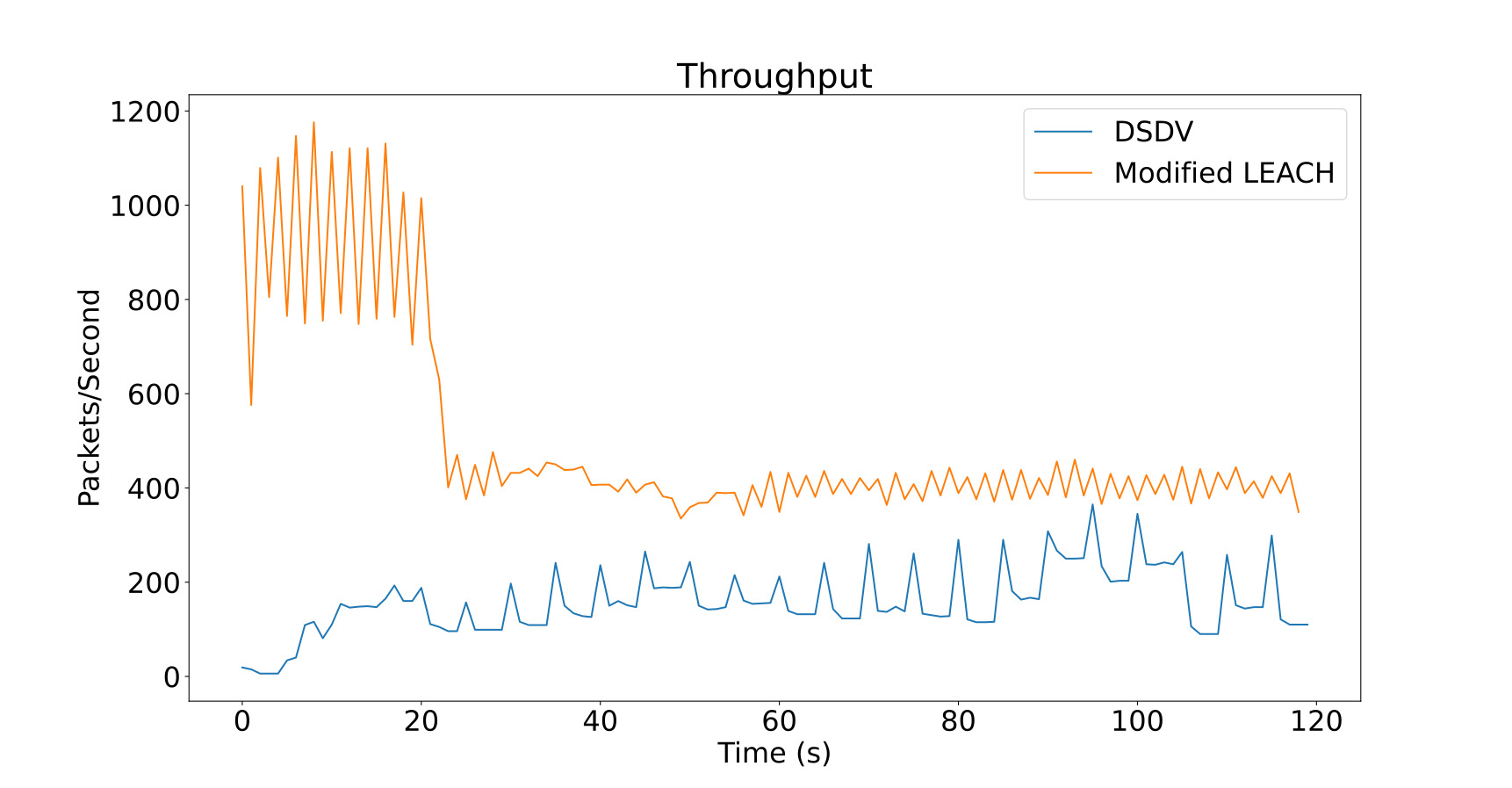}
    \caption{Throughput comparison of DSDV and MLEACH}
    \label{fig:throughput}
\end{figure}

Figure \ref{fig:total_energy} reveals a linear increase in the energy consumption as time passes. Each node in DSDV consumes a maximum of 0.819J compared to 0.836J for MLEACH to transmit data to the BS. 
Figure \ref{fig:throughput} shows the number of packets received by the BS per second. The drop in throughput occurs when the number of available CH reduces according to equation \ref{eqn:cluster_head_election}. While the MLEACH protocol consumes slightly more energy than DSDV, the protocol's throughput is vastly superior. The major drawback of DSDV is the constant updating of its routing tables resulting in more energy consumption. MLEACH can support an average of 400 packets per second in steady-state, and DSDV is only 200 packets per second while consuming similar amounts of energy. DSDV would require double the energy to match the throughput of MLEACH. This reduces the efficiency as compared to MLEACH. Thus, MLEACH is more energy-efficient and can last longer as compared to DSDV. 

\section{Conclusion}
\label{section:Conclusion}
 The flood is the most devastating natural disaster of all that ravages human settlement throughout the globe every year. In India alone, close to two thousand people lose their life due to the torrential floods that occur every year. We have proposed a model to optimize the rescue operation with the help of wireless sensor networks.  The proposed Wireless Sensor Network uses the custom MLEACH protocol that supports multi-hop routing and a novel cluster head election algorithm to make the network more power-efficient. The WSN nodes harbour essential sensors and transducers to help attract and locate the location of victims, emergency priority and flow of the flood and transmits relevant data to the base station to optimize the relief operation during floods. The nodes are deployed in an aerial vehicle with a centrifugal deployment system to ensure a wide coverage area. Sensor nodes are designed based on landing impact, flotation stability, ease of deployment, and cost-effectiveness to scale to the requirement. The proposed model is far better than other communication models of a sensor network in terms of throughput.  The model has been tested in NS3 simulation, resulting in greater throughput with the same power consumption compared to the DSDV protocol. This additional throughput can establish communication with the victims making the rescue process smooth and effective.

\section{Acknowledgment}
\label{section:Acknowledgment}
The authors are listed in ascending alphabetical order of names of the author who have contributed equally to this paper. We thank our guide, Dr Ujjwal Verma, for his constant support and guidance. Also, thanks to Mars Rover Manipal for supporting and sponsoring the research.  

\printbibliography

\end{document}